\input{aipcheck}

\documentclass[
    ,final            
  ]
  {aipproc}

\layoutstyle{6x9}

\usepackage{natbib}
\usepackage{latexsym}
\bibpunct{(}{)}{;}{a}{}{,}

\begin{document}

\title{Accreting millisecond pulsars: one on each hand}


\author{Manuel Linares}{
  address={Astronomical Institute ``Anton Pannekoek'', University of Amsterdam, Kruislaan 403, NL-1098 SJ Amsterdam, Netherlands. mail to: mlinares@science.uva.nl}
}

\author{Michiel van der Klis}{
}

\author{Rudy Wijnands}{
}

\begin{abstract}

We report on the X-ray aperiodic timing analysis of two accreting millisecond pulsars: XTE~J1807--294 and IGR~J00291+5934. On the one hand, we discovered in XTE~J1807--294 seven pairs of simultaneous kilohertz quasi-periodic oscillations (kHz QPOs) separated in frequency by nearly the spin frequency of the neutron star. This confirms the suspected dichotomy in the frequency separation of kHz QPOs: sometimes once and sometimes half the spin frequency. On the other hand, we found an extreme behavior in the power spectra of IGR~J00291+5934: very strong variability at very low frequencies. Namely, the fractional amplitude of the variability was $\sim$50\%, the highest value found so far in a neutron star system. 

\end{abstract}

\maketitle


\section{Introduction}

Weakly-magnetic neutron stars that accrete matter from low-mass companion stars form the $\sim$~150 neutron star low-mass X-ray binaries (NS-LMXBs) observed in our Galaxy since the early years of X-ray astronomy to date. Eight years ago the first of such systems showing X-ray millisecond pulsations was discovered by \citet[SAX~J1808.4--3658,][]{Wijnands98}, presenting the first prove of an accreting neutron star having both millisecond spin period and dynamically important magnetic field. At the moment of writing a total of seven such accreting millisecond pulsars (AMPs) have been found in faint, compact or ultracompact LMXBs, all of them transients \citep{Wijnands05}. Not only are they thought to be progenitors of radio millisecond pulsars but they also constitute an excellent test bed to study the accretion flow in NS-LMXBs and its interaction with the neutron star. 

The main open question concerning AMPs is what makes them X-ray pulsars. In general NS-LMXBs do not show millisecond pulsations, although many of them are now known to harbor a neutron star with millisecond spin period thanks to the oscillations seen during type I X-ray bursts (``burst oscillations''; see e.g. \citep{Strohmayer96, Chakrabarty03}). However, the seven members of the recently discovered and gradually growing familiy of AMPs show clear and strong pulsations during outburst. The typical pulsed fraction in AMPs ranges from 5 to 10~\%, whereas in non-pulsing NS-LMXBs the current upper limits are between 0.3 and 0.8~\% \citep{Vaughan94, Dib04}. As mentioned above, the pulsations observed in AMPs imply that the neutron star has a magnetic field strong enough ($\sim 10^8$-$10^9$~G) to channel the accretion flow and create thereby an asymmetric emission pattern (two hot spots in the simplest view). The question is then, why the rest of NS-LMXBs do not show pulsations? One option is that spin and magnetic axes are aligned in the non-pulsing systems and misaligned in AMPs, or that pulsations are smeared out in non-pulsing NS-LMXBs by an optically thick cloud surrounding the neutron star \citep{Titarchuk02}. A different physical mechanism was proposed by \citet{Cumming01} invoking screening of the magnetic field in ``classical'' NS-LMXBs by a time-averaged mass accretion rate higher than that of AMPs. In that case, however, it seems surprising that the spectral and timing properties of the low-luminosity NS-LMXBs (atoll sources) are so similar to those of AMPs. In other words, the accretion flows in AMPs and non-pulsing systems seem almost (see last Section) identical whereas in this scenario their magnetic field strengths are significantly different. If a dipolar magnetic field were important in AMPs and instead a quadrupolar or higher order magnetic field were to be dominant in atoll sources this would allow for similar magnetic field strengths affecting the accretion flow and would still be compatible with the lack of pulsations in atoll sources. A magnetic field dominated by high-order multipoles would create multiple emitting regions (see \citet{Krolik91}) whereas a pure quadrupolar magnetic field could give rise to a ``hot belt'' (see recent work by \citet{Long06}). This would in turn wash out pulsations, unless the emission were highly beamed. It is perhaps interesting to note that multipolar magnetic fields have been explored in the context of millisecond radio pulsars, the likely evolutionary descendants of AMPs, as possible causes of peculiar spin-down, pulse shape or of the observed X-ray emission \citep{Cheng03, Krolik91}.

One way to study accretion onto compact objects is to analyze the aperiodic variability in the X-ray flux coming from these sources, which tells us about processes occuring in the inner accretion flow \citep{vanderKlis95,vanderKlis06}. We performed such aperiodic timing analysis on two AMPs: XTE~J1807--294 and IGR~J00291+5934, and combined it with the spectral information obtained from X-ray colors. In the former we discovered pairs of kilohertz quasi-periodic oscillations (kHz QPOs; refs.) whose frequency separation was nearly equal to the spin frequency of the neutron star \citep{Linares05}. Surprisingly, in the latter we found instead a behavior atypical for NS-LMXBs and more similar to that observed in systems where the accreting compact object is a black hole candidate (black hole X-ray binaries, BH-XBs): very strong X-ray variability occurring at very low Fourier frequencies. We used in our work data collected by the proportional counter array (PCA) onboard the Rossi X-ray timing explorer ({\it RXTE}).

\section{XTE~J1807--294}

On February $13^{th}$, 2003, a new transient X-ray source was discovered in the Galactic bulge region and soon after that 190.6~Hz coherent pulsations were detected turning the new system, XTE~J1807--294, into the fourth discovered AMP \citep{Markwardt03c}. The orbital period was found to be $\sim$40 minutes \citep{Markwardt03a}, still the shortest among AMPs. The outburst was followed by {\it RXTE} during five months.

\begin{figure}[h]
  \includegraphics[width=.8\textwidth]{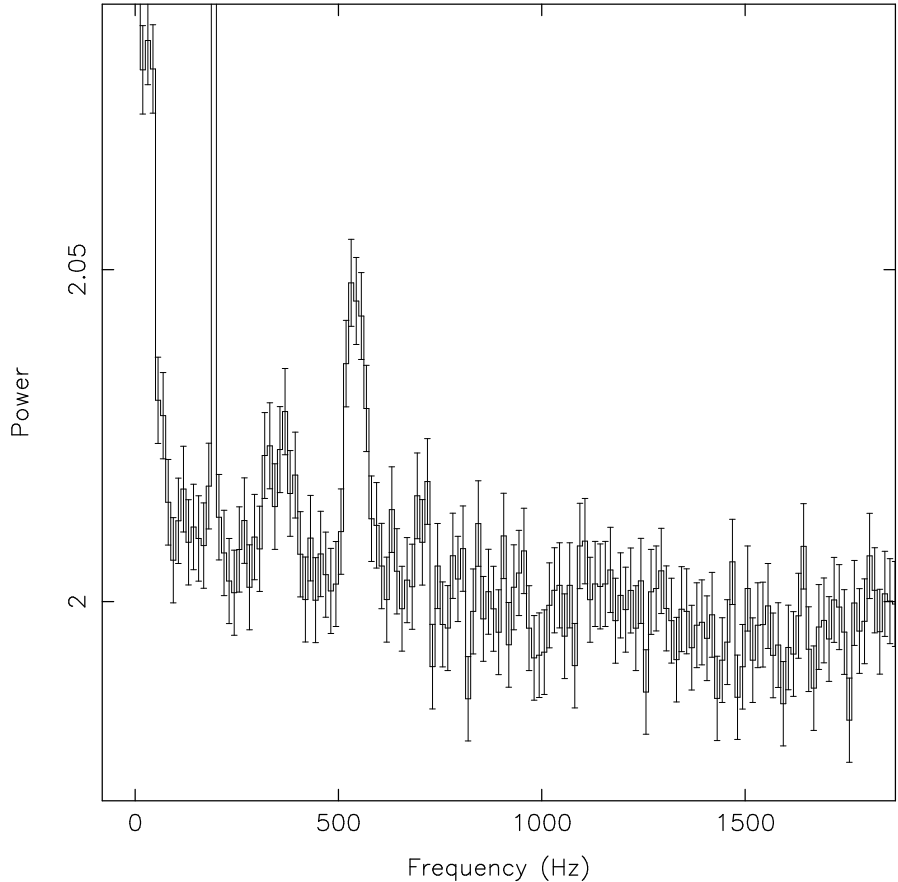}
  \caption{Twin kilohertz quasi-periodic oscillations (kHz QPOs) discovered in the X-ray flux of XTE~J1807--294. The pulse spike is visible near 200~Hz.}
\label{fig:khzqpos}
\end{figure}
\begin{figure}[h!!!]
 \includegraphics[width=.8\textwidth]{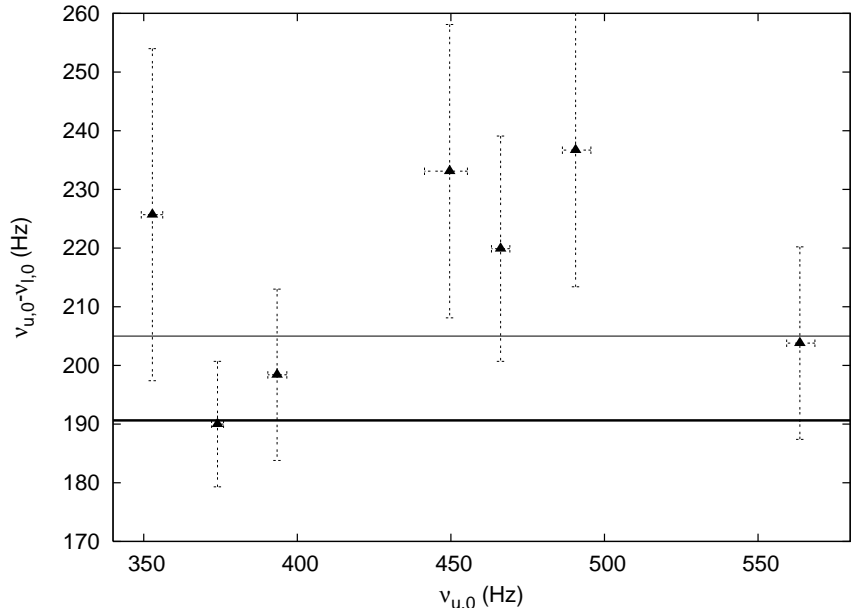}
  \caption{Frequency separation between the kHz QPOs detected during XTE~J1807--294's 2003 outburst versus the frequency of the upper kHz QPO. Errors are 1-sigma. The weighted average (thin line) and the spin frequency (thick line) are also shown.}
\label{fig:deltanu}
\end{figure}


kHz QPOs are (in most cases pairs of) variable-frequency narrow variability features observed in the power density spectra of about 20 NS-LMXBs. They are generally thought to come from the inner regions of the accretion disk and therefore constitute a potential probe of the extreme gravitational field surrounding the neutron star. Several models have been proposed for the kHz QPOs \citep{Miller98,Stella98,Abramowicz03}. Since the discovery by \citet{Wijnands03} of one pair of simultaneous kHz QPOs separated by {\bf half} the spin frequency ($\nu_{spin}$) in SAX~J1808.4--3658 a new set of models emerged trying to explain the intimate link between kHz QPOs and neutron star spin \citep{Wijnands03, Lamb03, Kluzniak04, Lee04}. We found seven pairs of twin kHz QPOs in XTE~J1807--294 separated instead by approximately {\bf once} the spin frequency (Figs.~\ref{fig:khzqpos} and \ref{fig:deltanu}; see \citet{Linares05} for further details). This confirms for the first time from a pulsating NS-LMXB that in some systems the frequency separation between kHz QPOs ($\Delta\nu$) is nearly equal to the spin frequency of the neutron star, and together with the results from SAX~J1808.4--3658 \citep{Wijnands03} clarifies the relation between $\Delta\nu$ and $\nu_{spin}$. According to the present data a model for kHz QPOs must explain why $\Delta\nu$ is rather constant and nearly equal to $\nu_{spin}/2$ (when $\nu_{spin}$$>$400~Hz, in the ``fast rotators'') or $\nu_{spin}$ (when $\nu_{spin}$$<$400~Hz, in the ``slow rotators''). This happens in {\it all} (eight) systems where both $\Delta\nu$ and $\nu_{spin}$ (via pulse or burst oscillations measurements) are known. However, a successful model for kHz QPOs should also adress the fact that in some sources $\Delta\nu$ is clearly not constant \citep{vanderKlis97,Boutloukos06}, being sometimes significantly above \citep{Jonker02} or below \citep{Migliari03} the spin frequency or half this value.

\section{IGR~J00291+5934}

On December $2^{nd}$, 2004, the sixth AMP was discovered: IGR~J00291+5934 \citep{Eckert04}. Coherent pulsations were found at a frequency of 598.9~Hz, modulated by the $\sim$2.5~hr orbital motion \citep{Markwardt04b,Markwardt04c}. Follow up observations of the outburst, which lasted about two weeks, were performed by {\it RXTE}.

\begin{figure}[h!!!]
  \includegraphics[width=1.0\textwidth]{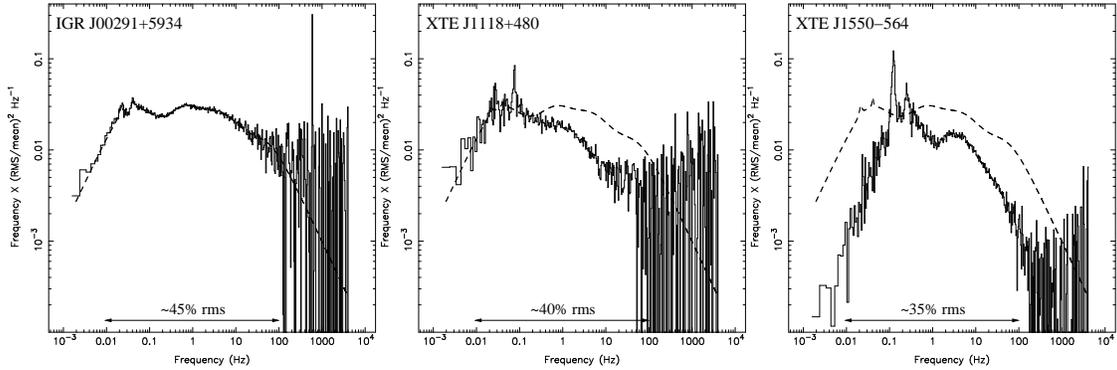}
  \caption{A power spectrum of IGR~J00291+5934 compared to those of two black hole low-mass X-ray binaries in their low state. Similarities up to $\sim$0.2~Hz are remarkable yet there are interesting differences above that. See \citet{Linares06} for further details and discussion.}
\label{fig:pscomp}
\end{figure}

Accretion onto compact objects in LMXBs takes place in/with different physical configurations giving rise to ``accretion states'', distinguished by different timing and spectral properties, through which systems transit. The main subdivision is that between low- (luminosity) -hard (spectra) and high- (luminosity) -soft (spectra) states, as observed in the X-ray band. AMPs are typically in low-hard states during their outbursts, very similar to the well studied island or extreme island states of non-pulsing low-luminosity NS-LMXBs \citep[i.e. the ``atoll'' sources;][]{Hasinger89, Straaten05, vanderKlis06}. Island and extreme island states have harder spectra, lower luminosities and stronger variability at lower frequencies than high-soft (banana) states. The case of IGR~J00291+5934 was exceptional as it showed even stronger variability, at frequencies lower than those typical of extreme island states. This was not a mere consequence of a very low mass accretion rate in the inner disk, as the X-ray luminosity measured during the outburst was not specially low, and could perhaps be related to a combination of dynamically important magnetic field and fast spin (see discussion in \citet{Linares06}). On the other hand, the X-ray variability we found is similar to that of black hole systems in the low-hard state (see Figure~\ref{fig:pscomp}). We measured the highest fractional rms amplitude observed so far in a NS-LMXB ($\sim$50~\%) and a broad-band flat-top noise with extremely low break frequencies ($\sim$0.04~Hz), similar or lower than the break frequencies measured in some black hole systems in the low state. We detected harmonically related QPOs on top of this break (at $\sim$0.02 and $\sim$0.04~Hz), a phenomenon also observed in low states of black hole systems (see Fig.~\ref{fig:pscomp}).

\section{Frequency correlations}

Characteristic frequencies of most of the different features present in the power spectra of NS- and BH-LMXBs are correlated with each other. Several studies in the last decade have demonstrated this \citep{WK99,PBK99,Straaten03} and have shown that there is some physical link between broad (noise) and narrow (QPO) components. When comparing the frequency correlations of some AMPs to those of previously studied atoll sources, \citet{Straaten05} found that SAX~J1808.4--3658 did not follow them but showed instead a clear shift in frequency. This shift was better explained by both the upper and the lower kHz QPOs having frequencies lower than the kHz QPOs seen in atoll sources, by a factor of 1.45$\pm$0.01. We found in XTE~J1807--294 a similar shift in the frequency correlations, by a factor of $\sim$1.59$\pm$0.03 \citep{Linares05}, and confirmed thereby this phenomenon that so far affects only AMPs. The exceptionally low frequencies present during IGR~J00291+5934's outburst allowed us to place a neutron star system in the lowest region of these frequency-frequency correlations (upper ``kilohertz QPO'' frequency below 100~Hz). These data, together with observations of the next outburst of IGR~J00291+5934, can provide a firm link between neutron star and black hole variability (see Fig. 8 in \citep{Linares06} and discussion therein).


\begin{theacknowledgments}

ML likes to thank the organizers of Cefal{\'u}-2006 for a stimulating and multicoloured conference. Thanks are also due to A. Gurkan, A. Patruno and P. Weltevrede for interesting discussions during the writing of this paper.

\end{theacknowledgments}

\clearpage



\bibliographystyle{aipprocl} 


\end{document}